\documentstyle[11pt]{article}

   \newcommand{\be}{\begin{equation}}
   \newcommand{\ee}{\end{equation}}
   \newcommand{\lb}{\label}
   
   \newcommand{\br}{{\bf r}}
   
   \newcommand{\bv}{{\bf v}}
   \newcommand{\bz}{{\bf 0}}
   \newcommand{\bA}{{\bf A}}
   
   \newcommand{\bl}{{\bf l}}
   \newcommand{\bj}{{\bf j}}
   \newcommand{\bs}{{\bf s}}
   \newcommand{\bD}{{\bf D}}
   \newcommand{\thetal}{\overline{\theta}}
   
   \newcommand{\vl}{\overline{{\bf v}}}
   
   \newcommand{\grad}{{\mbox{\boldmath $\nabla$}}}
   \newcommand{\bdot}{{\mbox{\boldmath $\cdot$}}}
   \newcommand{\bsigma}{{\mbox{\boldmath $\sigma$}}}

\begin{document}
   \title{Intermittency and Anomalous Scaling of Passive Scalars in Any Space
   Dimension}
   \author{Gregory L. Eyink\\{\em Department of Mathematics}\\
   {\em University of Arizona}\\{\em Tucson, AZ 85721}}
   \date{ }
   \maketitle
   \begin{abstract}
   We establish exact inequalities for the structure-function scaling exponents of
   a passively advected scalar in both the inertial-convective and
   viscous-convective ranges.
   These inequalities involve the scaling exponents of the velocity structure
   functions and, in a refined form, an intermittency exponent of the
   convective-range scalar flux.
   They are valid for 3D Navier-Stokes turbulence and satisfied within errors by
   present
   experimental data. The inequalities also hold for any ``synthetic'' turbulent
   velocity statistics with a finite correlation in time. We show that for
   time-correlation exponents
   of the velocity smaller than the ``local turnover'' exponent, the scalar
   spectral exponent is
   strictly less than that in Kraichnan's soluble ``rapid-change'' model with
   velocity
   delta-correlated in time. Our results include as a special case an
   exponent-inequality derived
   previously by Constantin \& Procaccia [Nonlinearity {\bf 7} 1045 (1994)], but
   with a more
   direct proof. The inequalities in their simplest form follow from a
   Kolmogorov-type
   relation for the turbulent passive scalar valid in each space dimension $d.$
   Our
   improved inequalities are based upon a rigorous version of the refined
   similarity
   hypothesis for passive scalars. These are compared with the relations implied
   by
   ``fusion rules'' hypothesized for scalar gradients.
   \end{abstract}

   \newpage

   \section{Introduction}

   Much progress has been made recently in the understanding of anomalous scaling
   for
   the problem of randomly advected scalars
   \cite{Kr94,KYC,FGLP,GK1,BGK,CFKL,CF,CFL,SS1,SS2}.
   The dynamical equation of the model is
   \be (\partial_t+\bv(\br,t)\bdot\grad_\br)\theta(\br,t)
		       =\kappa\bigtriangleup_\br\theta(\br,t)+f(\br,t) \lb{PS-EQ}
   \ee
   with $\theta(\br,t)$ the scalar field, $f(\br,t)$ a (stochastic or
   deterministic) source,
   and $\bv(\br,t)$ a random incompressible velocity field. The issue of interest
   is the presumed scaling law
   \be S_p(\ell)\sim \ell^{\zeta_p(\theta)} \lb{scal-law} \ee
   as $\ell\rightarrow 0$ for the scalar structure functions
   $S_p(\ell;\theta)\equiv \langle |\Delta_\ell\theta|^p\rangle$.
   For a high Reynolds number turbulence and for molecular diffusivity of the
   order of magnitude of the molecular
   viscosity, or greater, $\kappa \geq \nu,$ the dimensional theory of Obukhov
   \cite{Obu} and Corrsin \cite{Corr}
   implies that
   \be S_p(\ell)\sim (\chi/\varepsilon^{1/3})^{p/2}\ell^{p/3} \lb{Obu-Corr} \ee
   in which $\chi=\kappa\langle [\grad\theta]^2\rangle$ is the (mean) scalar
   dissipation and $\varepsilon$
   is the dissipation of kinetic energy. The scaling law Eq.(\ref{Obu-Corr}) is
   supposed to hold
   for $L\gg \ell\gg (\kappa^3/\varepsilon)^{1/4},$ where $L$ is the length scale
   of the scalar source,
   assumed less than the integral scale of velocity. This specifies the {\em
   inertial-convective range}.
   Thus, $\zeta_p(\theta)=p/3$ in the classical theory of this range. On the other
   hand, for $\kappa\ll \nu$
   there is another range, $L\gg \ell \gg (\kappa^3/\varepsilon)^{1/4},$ in which
   it is now assumed that $L$
   is at, or smaller than, the Kolmogorov dissipation scale
   $(\nu^3/\varepsilon)^{1/4}.$ Over this
   range, the so-called {\em viscous-convective range}, the theory of Batchelor
   \cite{Bat} proposes that
   \be S_p(\ell)\sim (\chi/\sigma)^{p/2}\log^{x_p}(\ell/L), \lb{Bat} \ee
   with $\sigma$ a mean shear strength. Here, $\zeta_p(\theta)=0$ formally for all
   $p.$ Whereas \cite{Obu,Corr}
   predicted $\zeta_p(\theta)$ to be a linear function of index $p,$ it is now
   generally expected that these
   exponents are some nontrivial concave functions of $p,$ i.e. that there is
   anomalous scaling \cite{AHGA}.

   The new work on this problem cited above mostly deals with a special model in
   which the Eulerian velocity
   field is zero-mean Gaussian, delta-correlated in time
   \be \langle v_i(\br,t) v_j(\br',t')\rangle =V_{ij}(\br-\br')\delta(t-t').
   \lb{vel-cov} \ee
   This so-called ``rapid-change model'' was first investigated by Kraichnan
   \cite{Kr68}, who
   observed that for this case the infamous closure problem is absent: the
   $N$th-order correlator
   of $\theta$ obeys equations depending only upon itself and {\em lower order}
   correlators.
   The recent analytical investigations explore particular limiting regimes: the
   case with space-dimension $d\gg 1$
   in \cite{CFKL,CF}  and the case with eddy-diffusivity exponent $0<\zeta\ll 1$
   in \cite{GK1,BGK}. The latter exponent is
   defined by the assumed scaling relation for the (Richardson) eddy-diffusivity
   tensor
   \begin{eqnarray}
   K_{ij}(\br) & \equiv &
		  {{1}\over{2}}\int_{-\infty}^t
   ds\,\,\langle[v_i(\br,t)-v_i(\bz,t)][v_j(\br,s)-v_j(\bz,s)]\rangle \cr
	      \,    & =      & V_{ij}(\bz)-V_{ij}(\br), \lb{Taylor}
   \end{eqnarray}
   that
   \be K_{ij}(\br)\sim D\cdot (d-1)\delta_{ij}r^\zeta+ D\cdot\zeta r^\zeta
				   \left(\delta_{ij}-{{r_i r_j}\over{r^2}}\right)
   \lb{K-scale} \ee
   for small $r.$ Note that Eq.(\ref{Taylor}) is an analogue for this model of
   Taylor's 1921
   exact formula for the eddy-diffusivity \cite{Tay} (which involves instead
   Lagrangian velocities).

   It is our purpose here to consider the problem with a finite time-correlation
   of the convecting velocity field. This includes the realistic case where the
   velocities are
   turbulent solutions of Navier-Stokes dynamics. In addition, our results apply
   to a model recently considered \cite{CFL} with Eulerian velocity field taken as
   a Gaussian
   with covariance obeying dynamical scaling
   \begin{eqnarray}
   V_{ij}(\br,t) & \equiv &
   {{1}\over{2}}\langle[v_i(\br,t)-v_i(\bz,t)][v_j(\br,0)-v_j(\bz,0)]\rangle \cr
	      \,    & =      & {{D
   r^\zeta}\over{\tau_r}}\left[\delta_{ij}g_{\|}\left({{t}\over{\tau_r}}\right)
		       +\left(\delta_{ij}-{{r_i
   r_j}\over{r^2}}\right)g_\perp\left({{t}\over{\tau_r}}\right)\right],
   \lb{dyn-scal}
   \end{eqnarray}
   with $\tau_r=\tau_L(r/L)^z.$  Our results in the finite correlation-time models
   shall be applicable
   to both the limiting regimes studied for the rapid-change model: $d\gg 1$ and
   $0<\zeta\ll 1.$ However, rather
   than asymptotic formulae for the scaling exponents, we shall derive exact
   inequalities. One interest of our
   results is that they point up some significant differences between the zero and
   finite correlation-time problems.

   Our simplest set of inequalities are based upon the following relation:
   \be \langle \Delta_\ell v_{\|}[\Delta_\ell\theta]^2\rangle =
   -{{4}\over{d}}\chi\cdot \ell, \lb{Yag} \ee
   This equation is valid for $L\gg \ell \gg \eta_D,$ where $L$ is the length
   scale of the scalar source and
   $\eta_D$ is a dissipation length, given by the Obukhov-Corrsin length
   $(\kappa^3/\varepsilon)^{1/4}$ \cite{Obu,Corr} for
   high Reynolds number Navier-Stokes turbulence and by $(\kappa/D)^{1/\zeta}$ for
   the model of \cite{CFL}
   (Eq.(\ref{dyn-scal})). Eq.(\ref{Yag}) is a relation analogous to that of
   Kolmogorov for the third-order velocity
   structure function \cite{K41-III} and it was proved for $d=3$ by Yaglom in 1949
   \cite{Yag}. It
   simply expresses the constancy of scalar flux over the convective range of
   scales. By direct
   application of the H\"{o}lder inequality, we will derive from this relation a
   basic set of inequalities relating
   the scaling exponents of $p$th structure functions of the scalar
   $\zeta_p(\theta)$ with those
   of the velocity $\zeta_p(v).$ The implications of these results will be
   discussed in Section 2.

   Furthermore, we shall derive an improved set of inequalities by means of a {\em
   refined similarity relation} (RSR)
   for the passive scalar. In a precise version stated below, the RSR we prove is
   \be X_\ell(\br)\sim {{\Delta_\ell
   v(\br)[\Delta_\ell\theta(\br)]^2}\over{\ell}}, \lb{RSR} \ee
   in which $X_\ell(\br)$ is a local scalar flux to scales $<\ell$ at space point
   $\br.$ (Cf. \cite{AHGA}).
   The corresponding inequalities will involve the {\em intermittency exponent}
   $\tau_p(X)$ of the scalar flux
   \be \langle |X_\ell|^p\rangle \sim \ell^{\tau_p(X)}, \lb{intmit} \ee
   which measures the increasing spatial spottiness of the flux as
   $\ell\rightarrow 0.$ These inequalities
   rigorously establish an intuitive fact: that convective-range intermittency of
   the scalar flux implies
   anomalous scaling of the scalar structure functions. These results are given in
   Section 3, along with
   some general discussion of refined similiarity hypotheses for passive scalars,
   including the relation
   to ``fusion rules'' proposed for scalar gradients.

   \section{Yaglom Relation Inequalities}

   We shall sketch here very concisely the proof of the Yaglom relation for any
   space dimension $d.$ See also
   \cite{SKS}, and, for more details, \cite{UF} and Appendix II of \cite{Ey1}. The
   first step is to define a (mean)
   ``physical space scalar flux'', via
   \be  X(\bl)\equiv \left.-{{1}\over{2}}{{d}\over{dt}}
   \langle\theta(\br,t)\theta(\br+\bl,t)\rangle\right|_{{\rm conv.},t=0}.
   \lb{scalflux} \ee
   The subscript ``conv.'' indicates that only the convective terms in
   Eq.(\ref{PS-EQ}) are used. A simple
   calculation using incompressibility and spatial homogeneity gives
   $X(\bl)=-{{1}\over{4}}\grad_\bl\bdot
   \langle\Delta_\bl\bv[\Delta_\bl\theta]^2\rangle.$ Assuming also spatial
   isotropy, the vector $\bA(\bl)=
   \langle\Delta_\bl\bv[\Delta_\bl\theta]^2\rangle$ is (for $d>2$) of the form
   $\bA(\bl)=A_\|(\ell)\hat{\bl},$ where
   $\hat{\bl}$ is the unit vector in the direction of $\bl.$ In the convective
   range of length-scales $\ell$
   with constant mean scalar flux, the Eq.(\ref{scalflux}) becomes
   $-4\chi=\grad_\bl\bdot\bA(\bl),$ or
   \be -4\chi= {{d-1}\over{\ell}}A_\|(\ell)+{{dA_\|}\over{d\ell}}(\ell).
   \lb{Apareq} \ee
   The only solution of this equation regular for $\ell\rightarrow 0$ is
   \be A_\|(\ell)= -{{4\chi}\over{d}}\ell. \lb{Apar} \ee
   This completes the derivation of the Eq.(\ref{Yag}). It is useful to remark
   here that the Yaglom
   relation does {\em not} hold in the Kraichnan ``rapid-change'' model. In fact,
   the lefthand side of
   the Eq.(\ref{Yag}) is not even well-defined in Kraichnan's model, since it is
   expressed by a product
   of $\Delta_\ell v_\|$ and $[\Delta_\ell\theta]^2$ {\em at a single instant}.
   However, with the delta-correlation
   in time, the velocity is a distribution-valued (generalized) process and the
   single-time values are not
   defined. The analogous result for Kraichnan's model is \cite{Kr68}
   \be S_2(\ell)\sim {{\chi}\over{D\cdot d^2}}\ell^{2-\zeta}, \lb{struc2} \ee
   which, like the Yaglom relation for the finite correlation-time velocity
   statistics, is an exact result
   in the delta-correlated model.

   A simple set of exponent-inequalities follow from the Yaglom relation as a
   straightforward application of
   the H\"{o}lder inequality. The inequalities for $\zeta_p(\theta)$ involve as
   well the exponents $\zeta_p(v)$
   of the (absolute) structure functions of velocity, $S_p(\ell;v)\equiv \langle
   |\Delta_\ell v|^p\rangle
   \sim \ell^{\zeta_p(v)}.$ They are simplest to state in terms of the exponents
   $\sigma_p(\theta)=
   \zeta_p(\theta)/p,\sigma_p(v)=\zeta_p(v)/p.$ By the Yaglom relation and the
   H\"{o}lder inequality,
   \begin{eqnarray}
   4\chi\ell & =    & \left|\langle \Delta_\ell v_{\|}[\Delta_\ell\theta]^2\rangle
   \right| \cr
	  \, & \leq & \langle |\Delta_\ell \bv|^q\rangle^{1/q}\cdot\langle
   |\Delta_\ell\theta|^p\rangle^{2/p}
					 \sim \ell^{2\sigma_p(\theta)+\sigma_q(v)}
   \lb{Hold}
   \end{eqnarray}
   for
   \be p\geq 2,\,\,\,\,{{2}\over{p}}+{{1}\over{q}}=1. \lb{condit} \ee
   As this relation must hold in the infinitely long convective range as
   $\ell\rightarrow 0,$ it follows that
   \be 2\sigma_p(\theta)+\sigma_q(v)\leq 1. \lb{YagHold1} \ee
   We have used the isotropic form of the Yaglom relation, but this is inessential
   (see \cite{Ey1}).
   The special case of Eq.(\ref{YagHold1}) for $p=\infty,\,\,q=1$ was previously
   derived by Constantin and
   Procaccia \cite{CP1} on the basis of estimates for Hausdorff dimensions of
   scalar level-sets: see Eq.(4.2) there.
   \footnote{The authors of \cite{CP1} were not specific about which order $p,q$
   of exponents were involved,
   but inspection of the proof shows that $p=\infty$ and $q=1$---in our
   notation---was used. Actually, most
   of their proof generalizes to general $p,\,\,q$ satisfying Eq.(\ref{condit}),
   except for the result on
   dimensions of level sets $E_\theta,$ $D_H(E_\theta)\leq
   d-\sigma_\infty(\theta),$ their Eq.(1.8), which
   used $p=\infty.$ Incidentally, we disagree with the conclusion of \cite{CP1}
   that the estimate is sharp, i.e.
   an equality. This claim in Section 4 is based on an opposite inequality,
   $2\sigma_\infty(\theta)+\sigma_1(v)\geq 1,$
   supposed to be derived in \cite{CP2}, which we dispute. In fact, examination of
   the argument of \cite{CP2}
   shows that an implicit assumption was made that a {\em spatially uniform}
   diffusive cutoff exists, below
   which length-scale the scalar graph is smooth. This assumption was used in the
   selection of $r_0$ in their
   inequality Eq.(4.23). To bound ${\rm vol}_d(G(B))$ from {\em below} requires
   that $G(B)$ is smooth on scales $\leq r_0$
   so that a uniform choice of $r_0$ may be made. It is possible, however, that
   there is a ``local fluctuating cutoff,''
   as postulated in some multifractal pictures: see \cite{FV}. In that case, the
   uniform choice of $r_0$ would fail and
   the result, as well as its proof, might break down.} Our very simple derivation
   here shows that it belongs to a family
   of inequalities which are a consequence just of the constancy of scalar flux.
   These inequalities express a
   complementarity between the regularity of the velocity and scalar fields: if
   the (Besov) regularity
   exponent of velocity $\sigma_q(v)$ is ``big'' then the corresponding scalar
   exponent $\sigma_p(\theta)$ must be ``small.''

   The basic inequality Eq.(\ref{YagHold1}) may also be written as
   \be \zeta_p(\theta)\leq {{p}\over{2}}(1-\sigma_q(v)) \lb{YagHold2} \ee
   for $p\geq 2,\,\,{{2}\over{p}}+{{1}\over{p}}=1.$ This relation may be
   considered under various special assumptions.
   If the scaling exponents of velocity $v$ are taken to be K41, i.e.
   $\sigma_q(v)={{1}\over{3}}$ for all $q\geq 1,$
   then it follows that
   \be \zeta_p(\theta)\leq {{p}\over{3}} \lb{ObCorIneq} \ee
   for $p\geq 2.$ Thus, the Obukhov-Corrsin predictions for the
   inertial-convective range appear as upper bounds.
   Likewise, if the velocity field is assumed smooth, or $\sigma_q(v)=1$ for all
   $q\geq 1$, then
   \be \sigma_p(\theta)\leq 0. \lb{BatcIneq} \ee
   The smoothness assumption would hold, for example, for the velocity field in
   the viscous dissipation range,
   and then the Batchelor exponents for the viscous-convection range appear as
   upper bounds. More generally, if it is
   assumed that
   $\sigma_q(v)=h$ for all $q\geq 2$, then
   \be \zeta_p(\theta)\leq {{p}\over{2}}(1-h). \lb{YagHoldGaus} \ee
   This assumption on the velocity scaling exponents corresponds to a
   ``monofractal'' field and would be true
   for convection by a Gaussian random velocity field.

   The last inequality has some interesting consequences for the model studied in
   \cite{CFL}. In that model
   the velocity field is spacetime Gaussian with covariance satisfying dynamical
   scaling, Eq.(\ref{dyn-scal}).
   Setting $t=0$ in that equation, it is easy to see that $2h=\zeta-z$ or
   \be \zeta=2h+z. \lb{HoldExp} \ee
   As pointed out above, $\zeta$ has roughly the interpretation of an
   eddy-diffusivity exponent analogous
   to the Richardson exponent $\zeta_{\rm R}={{4}\over{3}}$ \cite{Rch}. This is
   not entirely accurate since the exponent
   appears in the scaling law Eq.(\ref{dyn-scal}) hypothesized for {\em Eulerian
   velocities} in the model of \cite{CFL}.
   For $z<1$ Eq.(\ref{dyn-scal}) is not an accurate representation of Eulerian
   time correlations, which will
   then be dominated by convective sweeping. Nevertheless, keeping to this
   terminology, there is also for fixed
   ``eddy-diffusivity exponent'' $\zeta$ a complementarity between the magnitudes
   of velocity regularity exponent $h$
   and dynamical scaling exponent $z$: if one is big, the other is small. Of
   course, this is just due to the simple
   heuristic that eddy-diffusivity $K_\ell\sim v_\ell^2\tau_\ell.$  A similar
   relation should also hold for Navier-Stokes
   turbulence, i.e. $\zeta=\zeta_2(v)+z$, except that there the scaling exponent
   $z$ will correspond to {\em Lagrangian}
   time-correlations. It is only for Lagrangian time functions that the dynamical
   scaling can hold and, furthermore, the
   exact Taylor formula Eq.(\ref{K-scale}) involves such correlations. By means of
   Eq.(\ref{HoldExp}), the main inequality
   Eq.(\ref{YagHold2}) may be reexpressed as
   \be \zeta_p(\theta)\leq {{p}\over{4}}(z+\gamma), \lb{YagHoldGaus2} \ee
   where $\gamma\equiv 2-\zeta.$ In fact, Eq.(\ref{HoldExp}) states that
   \be 1-h={{z+\gamma}\over{2}}, \lb{HoldExp2} \ee
   so that it is a direct consequence. Eq.(\ref{YagHoldGaus2}) holds for all
   $p\geq 2$ in the model of \cite{CFL};
   furthermore, it will hold also for $p=2$ in Navier-Stokes turbulence, if
   $\zeta=\zeta_2(v)+z$ as expected.

   In \cite{CFL} an expansion about the ``rapid-change model'' was developed in
   the magnitude $\epsilon$ of the
   correlation-time. They employed a particular choice of dynamical exponent
   $z=\gamma.$ Their plausible physical argument
   for this choice ran as follows: the ``local turnover time'' of scalar eddies at
   scale $\ell$ is $t_\ell\sim \ell/v_\ell\sim
   \ell^{1-h}$ and this defines a ``local turnover value'' of the dynamical
   exponent $z=1-h.$  Note by using
   Eq.(\ref{HoldExp}) that this value is achieved precisely when $\zeta=1+h,$ or
   $\gamma=1-h.$
   In other words, the local turnover exponent is obtained when $z=\gamma.$ For
   $z>\gamma$ the velocity at vanishingly small
   scales changes randomly at a faster and faster rate relative to the evolution
   time of the scalar eddies. Hence, it is
   plausible that the predictions of the ``rapid-change'' model will hold in that
   case. On the other hand, for $z<\gamma,$ the
   realizations of the velocity field are selected randomly at a rate which goes
   to zero compared to the
   scalar cascade rate, i.e. the velocity field randomness is ``frozen in.'' We
   shall now observe that in the
   latter case of quenched randomness, or $z<\gamma,$ the scalar spectral exponent
   \be \zeta_2(\theta)< \gamma, \lb{SpecExp} \ee
   with strict inequality. This is direct from Eq.(\ref{YagHoldGaus2}). More
   generally, for $p\geq 2$
   \be \zeta_p(\theta)< {{p}\over{2}}\cdot\gamma. \lb{PExp} \ee
   Hence, all of the $\zeta_p$'s for $p\geq 2$ are {\em strictly} smaller than the
   ``classical'' values.
   This is not so surprising for the higher $p$-values, $p>2,$ since this is a
   familiar situation usually
   associated to ``intermittency.'' The present strict inequalities, including the
   unusual case $p=2$, have
   a different origin. They arise just from the condition of constant mean flux,
   which requires smaller
   $\zeta_p(\theta)$'s when $h$ is bigger.  However, $z$ smaller than the local
   turnover exponent at
   fixed value of ``eddy-diffusivity exponent'' $\zeta$ requires $h$ bigger than
   its classical value $1-\gamma.$
   Indeed, $1-\gamma< h< 1-z$ from Eq.(\ref{HoldExp2}) if $z<\gamma.$

   Another important comparison between the ``rapid-change'' model and the finite
   time-correlation cases
   arises from the dimension-dependence of the Yaglom relation. It is clear from
   Eq.(\ref{struc2}) for $S_2(\ell)$
   in the rapid-change model that it goes to a finite limit as
   $d\rightarrow\infty$ if and only if $D\propto D_0/d^2,$
   with $D_0$ fixed. In fact, it follows from the work of Chertkov et al. in
   \cite{CFKL} that with that choice of
   $d$-dependence of $D$ all scalar correlations have a nontrivial limit as
   $d\rightarrow\infty,$ and, in fact, correspond to
   the correlations of a Gaussian field. Note that this dependence of $D$ implies
   that each of the $i=1,...,d$ components
   of the velocity field $v_i(\ell)$ at scale $\ell$ has typical magnitude
   ${{\ell^{1-\gamma}}\over{\sqrt{d}}}$ and, likewise,
   each of the $d^2$ components of the strain tensor
   $\bsigma_\ell={{1}\over{2}}[(\grad\bv_\ell)+(\grad\bv_\ell)^\top]$
   are of order ${{\ell^{-\gamma}}\over{\sqrt{d}}}$. It may be argued on the basis
   of theory of random matrices that the
   typical strains along principal axes, the eigenvalues of $\bsigma_\ell$, are
   order $\ell^{-\gamma}$ as $d\rightarrow \infty$
   : see \cite{Kr74A,FFR}. This seems to be the correct scaling for a nontrivial
   limit, since the strain magnitudes gives the
   rate $\sim 1/t_\ell$ of scalar cascade. \footnote{Actually, the {\em principal
   eigenvalue} of $\sigma_\ell$ may be expected
   to determine the rate of scalar cascade: see \cite{Kr74A}. However, for random
   Wigner matrices with asymptotic semicircle
   distribution of eigenvalues, it is known also that the leading eigenvalue is
   within $O(1/d^{2/3})$ of the right edge of the
   spectrum.} The Yaglom relation shows that the matter is not so simple for the
   finite time-correlation situation.
   For that relation a scaling $v\sim 1/d$ is required to obtain a finite limit as
   $d\rightarrow \infty.$ This does not
   contradict the results of \cite{CFL}, since they take
   $\epsilon=\tau_\ell/t_\ell$ as the small parameter of their
   expansion. Working this through, one finds that this amounts to taking
   $v_\ell\sim {{D_0}\over{\sqrt{\epsilon d}}}
   \ell^{1-\gamma}$ and $\tau_\ell\sim \epsilon\cdot{{\ell^\gamma}\over{D_0}}$.
   For any $d$, this correctly recovers the
   delta-correlated model in the limit $\epsilon\rightarrow 0.$ In fact,
   $\epsilon$ is just the quantity denoted $\tau_*$
   in \cite{Kr68}. For the validity of the Yaglom relation it is therefore
   required that $\epsilon\sim d$ and that is
   clearly incompatible with the condition of \cite{CFL} that $\epsilon\ll 1$ at
   large $d$. \footnote{The proportionality
   of single-time velocity realizations to $\epsilon^{-1/2}$ here, as well as in
   Kraichnan's original 1968
   derivation, makes clear that such single-time values do not exist in the
   delta-correlated model obtained by the limit
   $\epsilon\rightarrow 0.$ Only integrals over some finite time-interval are
   well-defined, and the lefthand side
   of the Yaglom relation, for the idealized white-noise limit, is a meaningless
   expression.}

   \section{Refined Similarity Inequalities}

   We shall now derive inequalities which improve those from the Yaglom relation.
   The basic idea of the proof is a
   scaling relation between the {\em local scalar flux} variable $X_\ell(\br)$ and
   the difference variables of velocity
   and scalar at the same point $\br$, Eq.(\ref{RSR}). This is an analogue of the
   {\em refined similarity hypothesis}
   (RSH) in 3D which---in the version of Kraichnan \cite{Kr74B}---states that
   local energy flux scales as
   $\Pi_\ell(\br)\sim [\Delta_\ell v(\br)]^3/\ell$ in terms of the velocity
   difference at the same point. The proofs
   given below follow closely methods used in our discussion of the 3D RSH in
   \cite{Ey2} and the 2D RSH for vorticity
   scaling exponents in \cite{Ey1}. If we assume
   \be \langle |X_\ell|^p\rangle\sim \ell^{\tau_p(X)}, \lb{5} \ee
   then there follow heuristically from Eq.(\ref{RSR}) relations between the
   exponents $\zeta_p(\theta),\zeta_q(v)$ and
   $\tau_r(X).$ The exponents $\tau_p(X)$ measure the increasing spatial
   intermittency or ``spottiness'' of the scalar flux
   at decreasing length-scales. In fact, since $\langle X_\ell \rangle=\chi$ over
   the long convective interval of $\ell,$
   it may be expected that the (concave in $p$) exponent $\tau_p(X)$ is {\em
   negative} for monents $p>1$.
   The corresponding growth in moments of $X_\ell$ as $\ell\rightarrow 0$ reflects
   the increase in its fluctuations.
   As we shall establish below, the intermittency of the convective range scalar
   flux implies anomalous scaling of the
   scalar structure functions over that same interval. It is this connection
   between intermittency and anomalous
   scaling which is the essence of Kraichnan's RSH \cite{Kr74B}. After these
   results are derived as theorems below,
   we shall comment on the relationship with other refined similarity hypotheses
   for passive scalars recently proposed
   \cite{Hos,ZAH,SKS}. These latter hypotheses are motivated by the original
   Kolmogorov RSH \cite{K62} which
   involves space-averaged dissipation rather than local flux.

   We first must introduce an appropriate definition of the local scalar flux. It
   is most easily
   done using a smooth {\em filtering technique} to differentiate the large-scale
   and small-scale
   modes. This is the same method used in the large-eddy simulation (LES)
   modelling scheme and in our
   earlier discussion of the 3D case \cite{Ey2}. Here we apply the filter to the
   scalar equation
   Eq.(\ref{PS-EQ}). That is, we consider the ``large-scale scalar field'' defined
   as the convolution field
   $\thetal_\ell=G_\ell*\theta,$ with some suitable filter function $G_\ell.$ The
   resulting equation is
   \be
   \partial_t\thetal_\ell(\br,t)+\grad\bdot(\vl_\ell(\br,t)\thetal_\ell(\br,t)
   +\bj_\ell(\br,t))=0. \lb{7} \ee
   The large-scale velocity field is likewise defined by $\vl_\ell=G_\ell*\bv.$
   Note that
   $\bj_\ell\equiv \overline{(\bv\theta)}_\ell-\vl_\ell\thetal_\ell$ is a
   space-flux of the scalar induced by
   the turbulent convection (eddy diffusion). A main ingredient of our proofs is
   the following exact formula for
   this turbulent flux:
   \be \bj_\ell(\br,t)=[\Delta\theta(\br,t)\Delta\bv(\br,t)]_\ell-
			       [\Delta\theta(\br,t)]_\ell[\Delta\bv(\br,t)]_\ell.
   \lb{8} \ee
   Here $[f]_\ell=\int d^2\bs\,\,G_\ell(\bs)f(\bs)$ is the average over the
   separation-vector $\bs$
   in the difference-operator $\Delta_\bs$ with respect to the filter function
   $G_\ell(\bs).$ See \cite{Ey1,Ey2}.
   This relation shows that $\bj_\ell(\br)\sim \Delta_\ell\bv(\br)\cdot
   \Delta_\ell\theta(\br)$
   at each space-point $\br.$

   Recall that the {\em scalar-intensity} integral
   $K(t)={{1}\over{2}}\int_\Lambda\,\,\theta^2(t)$
   is formally conserved by the full dynamics. From the Eq.(\ref{7}) for the
   large-scale scalar field it is
   straightforward to derive by the standard methods of nonequilibrium
   thermodynamics a local balance equation
   for its large-scale intensity $K_\ell\equiv {{1}\over{2}}\thetal_\ell^2.$ It
   has the form
   \be \overline{D}_t K_\ell(\br,t)+\grad\bdot\bD_\ell(\br,t)=-X_\ell(\br,t).
   \lb{9} \ee
   Here, $\overline{D}_t$ represents $\partial_t+\vl_\ell(\br,t)\bdot\grad_\br$,
   the convective derivative
   associated to the large-scale velocity,
   \be \bD_\ell(\br,t)\equiv \thetal_\ell(\br,t)\bj_\ell(\br,t) \lb{10} \ee
   represents space-transport of the large-scale intensity by convective
   diffusion, and the {\em scalar flux}
   \be X_\ell(\br,t)\equiv -\grad\thetal_\ell(\br,t)\bdot\bj_\ell(\br,t) \lb{11}
   \ee
   represents the scalar transfer to the small-scale modes. In a homogeneous,
   stationary ensemble
   the lefthand side of Eq.(\ref{9}) has vanishing average. In a steady-state with
   constant mean flux $\chi$
   of scalar substance to high wavenumbers, the average $\langle
   X_\ell\rangle=\chi,$ a constant, for $\ell$ lying
   in the convective interval. Together with Eq.(\ref{8}), the formula
   Eq.(\ref{11}) for scalar flux shows that
   $X_\ell(\br)\sim \Delta_\ell\bv(\br)\cdot [\Delta_\ell\theta(\br)]^2/\ell,$
   which is the RSR, Eq.(\ref{RSR}).
   It is the exact equations Eqs.(\ref{8}),(\ref{11}) which are the precise form
   of our RSR, applicable even without
   assumptions of local isotropy or other statistical properties. They are
   essentially kinematic in nature, based
   only upon the conservation properties of the underlying dynamics.

   {}From these exact relations, there follow refinements of the previous
   exponent-inequalities. In fact,
   it follows from the (generalized) H\"{o}lder inequality that for
   \be {{2}\over{p}}+{{1}\over{q}}={{1}\over{r}}, \lb{condit2} \ee
   the ordering holds that
   \begin{eqnarray}
   \ell^{\tau_r(X)/r} & \sim & \langle |X_\ell|^r\rangle^{1/r} \cr
		   \, & \leq & {{\langle
   |\Delta_\ell\theta|^p\rangle^{2/p}\langle|\Delta_\ell\bv|^q\rangle^{1/q}}\over
					 {\ell}} \cr
		   \, & \sim & \ell^{2\sigma_p(\theta)+\sigma_q(v)-1}, \lb{deriv}
   \end{eqnarray}
   and thus,
   \be  2\sigma_p(\theta)+\sigma_q(v)\leq 1+{{\tau_r(X)}\over{r}}. \lb{RefIneq}
   \ee
   This is our main result on the scaling exponents. The derivation requires only
   the exact kinematical relations,
   Eqs.(\ref{8}),(\ref{11}), rather than the heuristic form of the RSR,
   Eq.(\ref{RSR}). For the details of
   this, see the Appendix of \cite{Ey2}. Note that $r=1$ recovers the previous
   inequalities,
   Eq.(\ref{YagHold1}), if $\tau_1(X)=0.$ However, as noted above, intermittency
   of scalar flux will
   imply that $\tau_r(X)<0$ for $r>1,$ and then the inequalities are sharpened.
   This is particularly easy
   to see for the case of a ``synthetic'' turbulent convection by a Gaussian
   random velocity field.
   For the Gaussian field (or any ``monofractal'' field) $\sigma_q(v)=h$ for all
   $q.$ Thus, taking
   $q\rightarrow \infty$ in the above inequality, one easily obtains $r=p/2$ and
   \be \zeta_p(\theta)\leq {{p}\over{2}}(1-h)+\tau_{{{p}\over{2}}}(X).
   \lb{RefIneq2} \ee
   Thus, $\zeta_p(\theta)$ will be {\em strictly less} than the ``classical
   exponent''
   $\zeta_p^{{\rm class}}={{p}\over{2}}(1-h)$ and the ``anomaly'' is exactly an
   intermittency
   exponent of the convective-range flux. This result implies that convection by a
   regular-scaling
   random velocity will nonetheless lead to anomalous scaling for the scalar it
   passively convects,
   if the associated flux variable develops strong fluctuations.

   It is worthwhile to make a comparison of these results with the other
   recently-proposed RSH for passive scalars
   \cite{Hos,ZAH,SKS}. While our formulation is motivated by the 1974
   ``revisionist'' RSH of Kraichnan, involving
   flux, the RSH explored by the above authors is an adaptation of that originally
   proposed by Kolmogorov \cite{K62}.
   That is, it is hypothesized that the random variables
   \be V_\theta(\br,\bl)\equiv
   \Delta_\bl\theta(\br){{(\ell\varepsilon_\ell(\br))^{1/6}}
		    \over{(\ell\chi_\ell(\br))^{1/2}}}, \lb{RSHP} \ee
   defined in terms of volume-averaged dissipations $\varepsilon_\ell(\br)$ and
   $\chi_\ell(\br)$ of velocity and
   scalar intensities, respectively, have conditional distributions given values
   $\varepsilon_\ell$ and
   $\chi_\ell$, which are independent of the local Reynolds number $Re_\ell$ and
   local P\'{e}clet number $Pe_\ell$
   when those are both large. That is, the variable $V_\theta$ is supposed to have
   a universal distribution in
   the inertial-convective range of $\ell.$ If this relation is combined with the
   original Kolmogorov RSH, then
   it is easy to infer likewise the existence of a random variable $W_\theta$,
   universal in the same sense, such that
   \be \Delta_\ell v\cdot[\Delta_\ell\theta]^2 \sim W_\theta\cdot\chi_\ell\ell.
   \lb{RSHP'} \ee
   See \cite{SKS,ZAH}. Given the Kolmogorov RSH for velocity-differences, this
   last relation is, in fact, equivalent
   to the RSH for passive scalars proposed in \cite{Hos,ZAH,SKS}. It provides a
   natural bridge between that RSH
   for passive scalars and the one established here. Since we have shown that
   $X_\ell\sim \Delta_\ell v\cdot
   [\Delta_\ell\theta]^2/\ell,$ the above relation may be more or less paraphrased
   as
   \be     X_\ell\sim W_\theta\cdot\chi_\ell. \lb{Xchi} \ee
   In other words, the ratio $X_\ell/\chi_\ell\equiv W_\theta$ is a random
   variable whose distributions
   conditioned on fixed $\varepsilon_\ell,\chi_\ell$ are universal in the
   inertial-convective interval of $\ell.$ Again, this
   is essentially just a reformulation of the RSH of \cite{Hos,ZAH}. If it holds,
   then a simple consequence is that
   \be   \langle |X_\ell|^p\rangle=\langle
   |W_\theta|^p|\varepsilon_\ell,\chi_\ell,\ell\rangle
			    \cdot \langle\chi_\ell^p\rangle, \lb{momrel} \ee
   and the coefficient $\langle
   |W_\theta|^p|\varepsilon_\ell,\chi_\ell,\ell\rangle$ is, in the
   inertial-convective
   range, just a constant factor $\langle |W_\theta|^p\rangle.$ Therefore, in
   particular, $\tau_p(X)=\tau_p(\chi)$
   for all $p,$ and the intermittency-exponents of the convective-range flux
   $X_\ell$ and the scalar dissipation,
   volume-averaged over the same length-scales, $\chi_\ell,$ are the same. In that
   case, all of the inequalities
   previously rigorously derived in terms of $\tau_p(X)$ hold also for
   $\tau_p(\chi).$ In terms of providing a
   theoretical foundation to the RSH for passive scalars, it may be easier to
   proceed by starting with Eq.(\ref{Xchi}).

   Another interesting comparison involves the {\em additive fusion rule} (AFR),
   which was proposed originally
   for the turbulent velocity-gradients \cite{Ey3,LL}. Recently, the
   straightforward extension of these rules
   to the scalar-gradients has received some analytical support in Kraichnan's
   ``rapid-change'' model \cite{FGLP,BGK,CFKL,CF}.
   As we now explain, it happens that in this model the AFR and the RSH lead to
   identical relations between scaling
   exponents. The AFR states, in a schematic form, that
   \be [(\nabla\theta)^{p_1}]\cdots[(\nabla\theta)^{p_n}]\sim
   [(\nabla\theta)^{p_1+\cdots+p_n}]. \lb{AFR} \ee
   We ignore for the sake of this argument the vector character of the
   scalar-gradient which, properly, should
   be taken into account (Cf. \cite{Ey3}). The quantities $[(\nabla\theta)^p]$ are
   so-called ``renormalized
   composite variables.'' This means simply that one defines them as the limit of
   $p$th-powers of scalar-gradients
   at the same space-point, in the model with $\eta_D>0$, but multiplicatively
   renormalized by an appropriate power
   of $\eta_D.$ After nondimensionalizing the variable (according to its
   ``canonical'' or engineering
   dimension) there may still be required a factor $Z(\eta_D)\sim
   \left({{\eta_D}\over{L}}\right)^{-x_p}$
   to make the correlations finite in the limit $\eta_D\rightarrow 0.$ The
   exponent $x_p$ is the so-called
   {\em anomalous scaling dimension} of the variable $[(\nabla\theta)^p].$
   Observe, in this context, that the scalar
   dissipation $\chi(\br)=\kappa |\grad\theta(\br)|^2$---when divided by its mean
   $\bar{\chi}$---is nothing more than
   $[(\nabla\theta)^2].$  The precise meaning of the schematic result
   Eq.(\ref{AFR}) is that, inserted in arbitrary
   correlators at separated points,
   \begin{eqnarray}
   \, & &
   [(\nabla\theta)^{p_1}](\lambda\cdot\br_1)\cdots[(\nabla\theta)^{p_n}]
						     (\lambda\cdot\br_n) \cr
   \, & & \,\,\,\,\,\,\,\,\,\,\,\,\,\,\,\,\,\,\,\,\,\,\,\,\,\,\,\,
						     \,\,\,\,\,\,\,\,
	  \sim \lambda^{x_{p_1+\cdots+p_n}-x_{p_1}-\cdots-x_{p_n}}
		[(\nabla\theta)^{p_1+\cdots+p_n}](\bz), \lb{OPE}
   \end{eqnarray}
   in the limit as $\lambda\rightarrow 0.$ It is easy to show, as in \cite{Ey3},
   that the above AFR leads to
   a ``multiscaling law'' for scalar structure functions (now {\em without}
   absolute values ) of the form
   \be \langle [\Delta_\ell\theta]^p\rangle\sim \ell^{(1-x_1)p+x_p}. \lb{multisc}
   \ee
   Moreover, if the short-distance expansion is applied to the moments of the
   volume-averaged dissipation, it is
   immediately obtained that
   \be \langle \chi_\ell^p\rangle\sim \ell^{x_{2p}-px_2}. \lb{dispsc} \ee
   In Kraichnan's model, $x_1=1-(\gamma/2)$ and $x_2=0.$  Thus, Eq.(\ref{multisc})
   leads to
   $\zeta_p(\theta)=p(\gamma/2)+x_{p}$ and Eq.(\ref{dispsc}) yields
   $\tau_p(\chi)=x_{2p}.$ Hence, they together give
   \be \zeta_p(\theta)=p\cdot{{\gamma}\over{2}}+\tau_{{{p}\over{2}}}(\chi).
   \lb{AFR-RSH} \ee
   This should be compared with the result of RSH for the model,
   Eq.(\ref{RefIneq2}) [taken as an equality
   and with $\tau_p(\chi)$ replacing $\tau_p(X).$] Clearly, they are the same. It
   was already shown in \cite{CF}
   that the RSH holds in the white-noise model. Our point here is that the RSH is
   a consequence just of the AFR.
   \footnote{ This happy situation does not, however, hold for the original AFR
   applied to velocity-gradients,
   nor even necessarily for scalar-gradients in true turbulence. In the case of
   velocity-gradients, the
   application of the AFR analogous to the above leads to the relation \cite{Ey3}
   \be \zeta_p(v)=p\cdot {{\zeta_2(v)}\over{2}}+\tau_{{{p}\over{2}}}(\varepsilon),
   \lb{AFR-vel} \ee
   whereas the RSH leads instead to
   \be \zeta_p(v)= {{p}\over{3}}+\tau_{{{p}\over{3}}}(\varepsilon). \lb{RSH-vel}
   \ee
   These are not equivalent and the RSH result seems to be in better agreement
   with the experimental
   data (Sreenivasan, private communication, 1995). The failure of the AFR in this
   context could be related
   to the existence of a hierarchy of viscous cutoffs for the velocity field
   \cite{FV,Ey3}, or, possibly,
   to a naive disregard of the tensorial character of products of
   velocity-gradients.}

   One use of our exact inequalities is as a check on experimental data for
   scaling exponents, since these
   employ a variety of assumptions and approximations (Taylor hypothesis,
   one-dimensional surrogates, etc.)
   For that purpose, we may cite the experimental results of \cite{AHGA}
   \be \zeta_2(\theta)\approx 0.65,\,\,\,\zeta_3(\theta)\approx
   0.82,\,\,\,\zeta_4(\theta)\approx 0.95 \lb{expPS} \ee
   for the scalar exponents, and \cite{HvW}
   \be \sigma_2(v)\approx 0.355,\,\,\,\sigma_3(v)=0.333,\,\,\,\sigma_\infty(v)\leq
   \sigma_{17}(v)\approx 0.211 \lb{expV} \ee
   for the velocity exponents. We shall here assume
   $\sigma_{17}(v)\approx\sigma_\infty(v)$ since the graph of
   $\zeta_p(v)$ is close to linear for $p$ of order 20. In that case, if we make
   comparison, for simplicity,
   with the Yaglom inequalities, we find that
   \be 0.65\approx \zeta_2(\theta)\leq 1-\sigma_\infty(v)\approx 0.78 \lb{EQ1} \ee
   \be 0.82\approx \zeta_3(\theta)\leq {{3}\over{2}}(1-\sigma_3(v)) \approx 1.00
   \lb{EQ2} \ee
   \be 0.95\approx \zeta_4(\theta)\leq 2(1-\sigma_2(v))\approx 1.29 \lb{EQ3} \ee
   All of these inequalities are well-satisfied by the data. The fact that there
   is a considerable margin
   between the upper and lower limits is also consistent with an intermittency
   correction from the scalar
   flux. Determination of the latter from DNS or, experimentally, from the
   surrogate $\Delta_\ell v
   [\Delta_\ell\theta]^2/\ell,$ would be of interest.

   \noindent {\bf Acknowledgements.} I would like to thank all the participants of
   the 1995 CNLS summer workshop on
   fully-developed turbulence, where most of these results were originally
   presented, for their comments and questions,
   but especially R. H. Kraichnan and K. R. Sreenivasan. I also wish to thank G.
   Falkovich for very helpful
   discussions regarding his work.


\begin{thebibliography}{99}
   \bibitem[Kr94]{Kr94}R. H. Kraichnan, ``Anomalous scaling of a randomly advected
   passive scalar,'' Phys. Rev. Lett.
		       {\bf 72} 1016 (1994).
   \bibitem[KYC]{KYC}R. H. Kraichnan, V. Yakhot, and S. Chen, ``Scaling relations
   for a randomly advected passive
		     scalar field,'' {\bf 75} 240 (1995).
   \bibitem[FGLP]{FGLP}A. L. Fairhall, O. Gat, V. L'vov, and I. Procaccia,
   ``Anomalous scaling in a model of
		       passive scalar advection: exact results,'' Phys. Rev. E, in
   press (1996).
   \bibitem[GK1]{GK1}K. Gawedzki and A. Kupiainen, ``Anomalous scaling of the
   passive scalar,'' Phys. Rev. Lett.
		     {\bf 75} 3834 (1995).
   \bibitem[BGK]{BGK}D. Bernard, K. Gawedzki and A. Kupiainen, ``Anomalous scaling
   in the N-point functions of
		     passive scalar,'' preprint (1996), {\em chao-dyn/9601018}.
   \bibitem[CFKL]{CFKL}M. Chertkov, G. Falkovich, I. Kolokolov and V. Lebedev,
   ``Normal and anomalous scaling
		       of the fourth-order correlation function of a randomly
   advected scalar,'' Phys. Rev. E
		       {\bf 52} 4924 (1995).
   \bibitem[CF]{CF}M. Chertkov and G. Falkovich, ``Anomalous scaling exponents for
   a white-advected passive
		   scalar,'' submitted to Phys. Rev. Lett. (1996), {\em
   chao-dyn/9509007}.
   \bibitem[SS1]{SS1}B. I. Shraiman and E. D. Siggia, ``Lagrangian path integrals
   and fluctuations in random
		     flow,'' Phys. Rev. E {\bf 49} 2912 (1993).
   \bibitem[SS2]{SS2}B. I. Shraiman and E. D. Siggia, ``Anomalous scaling of a
   passive scalar in turbulent
		     flow,'' C. R. Acad. Sci. Paris, t.{\bf 321}, S\'{e}rie IIb,
   279 (1995).
   \bibitem[Obu]{Obu}A. M. Obukhov, ``Structure of the temperature field in a
   turbulent flow,''
		     Azv. Akad. Nauk. SSSR, Geogr. i Geofiz. {\bf 13} 58 (1949).
   \bibitem[Corr]{Corr}S. Corrsin, ``On the spectrum of isotropic temperature
   fluctuations in an isotropic
			turbulence,'' J. Appl. Phys. {\bf 22} 469-473 (1951).
   \bibitem[Bat]{Bat}G. K. Batchelor, ``Small-scale variation of convected
   quantities like temperature
		     in turbulent fluid,'' J. Fluid. Mech. {\bf 5} 113-133 (1959).
   \bibitem[AHGA]{AHGA}R. A. Antonia et al., ``Temperature structure functions in
   turbulent shear flows,''
		       Phys. Rev. A {\bf 30} 2704 (1984).
   \bibitem[Kr68]{Kr68}R. H. Kraichnan, ``Small-scale structure of a scalar field
   convected by turbulence,''
		       Phys. Fluids {\bf 11} 945 (1968).
   \bibitem[Tay]{Tay}G. I. Taylor, ``Diffusion by continuous movements,'' Proc.
   Lond. Math. Soc. (2)
		     {\bf 20} 196-211 (1921)
   \bibitem[CFL]{CFL}M. Chertkov, G. Falkovich, and V. Lebedev, ``Non-universality
   of the scaling exponents
		     of a passive scalar convected by a random flow,'' submitted
   to Phys. Rev. Lett.
		     (1996), {\em chao-dyn/9601016}.
   \bibitem[K41-III]{K41-III}A. N. Kolmogorov, ``Energy dissipation in locally
   isotropic turbulence,''
		     Dokl. Akad. Nauk. SSSR {\bf 32} 19-21 (1941).
   \bibitem[Yag]{Yag}A. M. Yaglom, ``On the local structure of a temperature field
   in a turbulent
		     flow,'' Dokl. Akad. Nauk. SSSR {\bf 69} 743 (1949).
   \bibitem[SKS]{SKS}G. Stolovitzky, P. Kailasnath, and K. R. Sreenivasan,
   ``Refined similarity hypotheses for passive
		     scalars mixed by turbulence,'' J. Fluid Mech. {\bf 297} 275
   (1995).
   \bibitem[UF]{UF}U. Frisch, {\em Turbulence: The Legacy of A. N. Kolmogorov}.
   (Cambridge U. Press,
		   Cambridge, 1995).
   \bibitem[Ey1]{Ey1}G. L. Eyink, ``Exact results on stationary turbulence in 2D:
   consequences of vorticity
		     conservation,'' Physica D, to appear (1996).
   \bibitem[CP1]{CP1}P. Constantin and I. Procaccia, ``The geometry of turbulent
   advection: sharp estimates
		   for the dimension of level sets,'' Nonlinearity {\bf 7} 1045
   (1994).
   \bibitem[CP2]{CP2}P. Constantin and I. Procaccia, ``Scaling in fluid
   turbulence: a geometric theory,''
		     Phys. Rev. E {\bf 47} 3307 (1993)
   \bibitem[FV]{FV}U. Frisch and M. Vergassola, ``A prediction of the multifractal
   model: the intermediate
		   dissipation range,'' Europhys. Lett. {\bf 14} 439 (1991).
   \bibitem[Rch]{Rch}L. F. Richardson, ``Atmospheric diffusion shown on a
   distance-neighbor graph,''
		     Proc. Roy. Soc. Lond. A {\bf 110} 709 (1926).
   \bibitem[Kr74A]{Kr74A}R. H. Kraichnan, ``Convection of a passive scalar by a
   quasi-uniform random
		       straining field,'' J. Fluid Mech. {\bf 64} 737 (1974).
   \bibitem[FFR]{FFR}J.-D. Fournier, U. Frisch, and H. A. Rose,
   ``Infinite-dimensional turbulence,''
		     J. Phys. A: Math. Gen. {\bf 11} 187 (1978).
   \bibitem[Ey2]{Ey2}G. L. Eyink, ``Local energy flux and the refined similarity
   hypothesis,'' J. Stat. Phys.
		    {\bf 78} 335 (1995).
   \bibitem[Kr74B]{Kr74B}R. H. Kraichnan, ``On Kolmogorov's inertial-range
   theories,'' J. Fluid Mech. {\bf 62} 305 (1974).
   \bibitem[Hos]{Hos}I. Hosokawa, ``Statistics of velocity increment in turbulence
   predicted from the Kolmogorov
		       refined similarity hypotheses,'' J. Phys. Soc. Jap. {\bf
   62} 10 (1993)
   \bibitem[ZAH]{ZAH}Y. Zhu, R. A. Antonia, and I. Hosokawa, ``Refined similarity
   hypotheses for turbulent velocity
		     and temperature fields,'' Phys. Fluids {\bf 7} 1637 (1995).
   \bibitem[K62]{K62}A. N. Kolmogorov, ``A refinement of previous hypotheses
   concerning the local structure of
		     turbulence in a viscous incompressible fluid at high Reynolds
   number,'' J. Fluid Mech. {\bf 13}
		     82 (1962).
   \bibitem[HvW]{HvW}J. Herweijer and W. van der Water, ``Universal shape of
   scaling functions in turbulence,''
		     Phys. Rev. Lett. {\bf 74} 4651 (1995).
   \bibitem[Ey3]{Ey3}G. L. Eyink, ``Lagrangian field theory, multifractals, and
   universal scaling in turbulence,''
		     Phys. Lett. A {\bf 172} 355 (1993).
   \bibitem[LL]{LL}V. V. Lebedev and V. S. L'vov, ``Scaling of correlation
   functions of velocity gradients in
		   hydrodynamic turbulence,'' JETP Lett. {\bf 59} 577 (1994).
   \end{thebibliography}
   \end{document}